# Detecting gasoline in small public places based on wireless sensor network

Rui Tian

Beijing Engineering Research Center for IoT Software and Systems, Beijing University of Technology, Beijing 100024, China

**ABSTRACT**

Fire accidents often cause unpredictable catastrophic losses. At present, existing fire prevention measures in public places are mostly based on the emergency treatments after the fire, which have limited protection capability when the fire spreads rapidly, especially for the flammable liquid explosion accident. Based on the gas sensor network, this paper proposes a detection framework as well as detail technologies to detect flammable liquid existing in small spaces. We propose to use sensor network to detect the flammable liquids through monitoring the concentrations of the target liquid vapor diffused in the air. Experiment results show that, the proposed surveillant system can detect the gasoline components in small space with high sensitivity while maintaining very low false detection rates to external interferences.

## 1 Introduction

Among common public safety accidents, fire disasters always lead to catastrophic consequences. Recently, the development of information technology is promoting city security emergency system to become a complete and powerful network. The IoT (Internet of things) technology has been widely used in lots of indoor places for security surveillances, such as home, office buildings, parks, production plants, warehouses and so on. These IoT systems are generally based on intrusion detection and video surveillance technologies [1], and can only detect simple malicious activities. For more complex fire prevention in public places, traditional means detect fire through discovering flame or combustion products in the environment, which will significantly reduce the response time to deal with the incident, and leads poor protective effect in small indoor spaces, such as office buildings, small meeting places, and motor coaches and so on.

In order to meet the fire prevention requirements for small public places, this paper proposes detection mechanism for flammable and explosive liquids detection. At present, in various industries, there are many monitoring systems and sensing technologies for such detection requirements. Such as liquid component detection based on infrared or ultrasonic absorption, measurement of gas production based on high-precision combustible gas meters. However, these methods generally require specialized environments, and also often rely on expensive equipment, which causes high system construction costs. However for fire preventing in common small public areas, it is impossible to customize surveillance environment for each scene. This paper proposes the fire detection mechanism in small space based on sensor network, which has low implementation cost as well as high scalability.

The contributions of this paper can be concluded into four parts: (1) In this paper, extensive experiments have been carried out to verify the feasibility of detecting gasoline volatiles using commercial combustible gas sensor probes; (2) Based on the experimental data, this paper proposes a judgment logic of anomalous gasoline gas diffusion based on temporal and spatial correlation analysis; (3) This paper presents a framework for the detection of gasoline volatile and data processing based on sensor networks; (4) Extensive experiments have been carried out to verify the effectiveness of the detection system.

This rest of this paper is organized as follows: Section 2 puts forward the design motivations of this paper. Section 3

introduces our sensor test experiments and our findings from the experiment results. Section 4 introduces the gasoline volatile detection logic adopted in this paper. Section 5 introduces the detailed mechanisms we adopt under the framework of gas detection. Section 6 verifies the detection performance of the system through real experiments. Section 7 concludes the whole paper.

## 2 Related work and Motivations

At present, much work has been done in automatic fire prevention, of which the most common applications include residential area fire prevention, forest fire prevention, mine fire prevention and so on [2]. For the residential area fire preventions, most work focuses on reducing the false alarm rate and at the same promoting the alarm rate of the system through improving the detecting techniques and data processing methods [3]. For the forest fire preventions, it mostly concerns that how to implement the system in the forest scenarios, and how to make the systems work reliably [4]. Mine fire preventions focus on building a system for complex mine topographies, requiring high system security, complete sensing coverage, and fast alarming ability to the fire [5]. This paper targets at fire detection for small public places, which has the similar design goals as residential fire prevention.

The work on automatic fire prevention for residential areas started very early. Since the 1990s, a series of research results have been published, such as selecting specific sensor sets to form a sensor array, designing a self-learning electronic nose to realize fire signs [6]. With the rapid development of wireless sensor network (WSN) technology in recent years, a considerable part of the work began to design networked automatic fire detection systems. As described in [7], an early fire detection system was developed, which was suitable for fire prevention in open spaces such as rural areas and urban areas. They incorporated temperature sensors and maximum likelihood algorithm to fuse sensory information. In [8], a WSN system for preventing fire accident on the running train was proposed. They monitored the temperature of the coaches to determine the fire. When the ambient temperature exceeds the critical temperature, all the drivers and passengers will be alarmed, the drivers can then manually stop the train and open water sprinkles all over the coaches. In [9], a fire-alarming system for indoor environment was proposed, which is capable of assisting firefighting activities including fire-alarming, fire-rescuing and firefighter orientation. Since fire usually spread quickly in small indoor environment, fire alarm based on detecting combustion products in the air will largely shorten the emergency response time. Focusing on detecting the flammable and explosive liquid in small public places, this paper tries to trigger fire alarm through detecting fire conditions, and is an early warning system.

Combustible gas monitoring is common in petrochemical's production and transportation [10]. In these applications, gas composition measurements usually need to follow standard operating procedures. In addition, there is also mature industrial chain in this field. There are a large number of high-precision gas sensors which can accurately measure the concentration of certain type gas components in the air. However, such sensors usually cost high and only have limited detection ranges. They also often occupy large installation spaces. On the other hand, to make these sensors work, usually a complete set of calibration routines is required, such as calibrating the sensor's measurement data by venting a volume of standard sample gas (such as hydrogen, isobutene, etc.) in a standard closed space. However, if the above calibration operation is carried out in a public place, the corresponding maintenance cost is unbearable, which will greatly hamper the actual system implementations.

On the other hand, in the common petrochemical applications, combustible gas detections usually concern whether or not the specific gas concentration exceeds critical threshold. However for combustible gas detection in public places, it is more important to judge the presence of combustible gas in monitoring places. Currently commercial gas sensors are designed to measure gas concentrations, and they are manufactured with preset detection ranges. In paper [11], the authors measured the concentration of gasoline volatiles at a distance of 3m from the discharge port in the gas station, and find that the gasoline vapor concentration fluctuates generally from 5ppm to 50ppm, which lies far below the lower bond of preset working range of most commercial combustible gas sensors. Whereas at the same time, through experiments, we luckily found that, most gas sensors can still detect gas diffusions even when the concentrations are very low and cannot be accurately measured. So we think that, it is feasible to use the commercial gas sensors to discover abnormal gas diffusions in small public areas by extracting and analyzing abnormal changes in the measurements of the sensors.

Aiming at identifying and quantifying the gas components, there are also sensor array-based detecting approaches. By matching the measured values of different types of sensors to unknown gases with the sample data set, the type and concentration of target gas in current environment can be determined. This type of method relies on a large number of sample trainings.

In general, gas detection in a small confined space has following characteristics: 1) Gas sensors operate in low concentration environments. The combustible gas sensor of the fire protection system operates below the preset linear working range of most commercial gas sensors. 2) Different gas selectivity requirement. Fire detection based on combustible gas sniffing is broad-spectrum sensing and does not target identified gas species. 3) Gas sensor is difficult to calibrate. Due to the variety of small public places, sensor calibration becomes very difficult. 4) Vulnerable to external environmental interference. External environmental vibration, and air flow, will affect the sensor's working condition, disturb measurement results.

Considering above characteristics, we propose the design objectives of this paper as: 1) High abnormal event reporting rate and very low abnormal event false alarm rate; 2) The ability to tolerate a variety of external disturbances such as airflow, temperature and humidity fluctuation, crowd movements in the monitoring spaces; 3) Do not need frequent calibration and maintenance, easy to deploy, and has a relative long working life time.

## 3 Sensor testing experiments

### 3.1 Sensor Selection

Common commercial combustible gas sensors include semiconductor, electrochemical, catalytic combustion, and infrared optical gas sensors and so on. According to different working principle, they have respective advantages and disadvantages, as well as different target applications. Among above sensors, the infrared optical sensor has the highest detection accuracy, but also has strong gas selectivity. This paper focuses on the detection of gasoline vapor. Since the gas composition of gasoline is very complex, and commercial infrared sensors do not have a specific model for gasoline vapor detection. If we choose other organic gas sensors instead, it is very likely to cause missed detection. For above other type of sensors, electrochemical sensors are rarely available for flammable and explosive gas detection products, catalytic combustion sensors typically have higher energy consumption. So in this paper, we choose the semiconductor combustible gas sensors, which have abandon product line and are widely used, to design the anomaly detection system.

We have selected several combustible gas sensors for testing, including models MQ6, MS6100, TGS813 and TGS2602. To verify these sensors' ability of detecting the low concentration of gasoline vapor, we conducted the following experiment.

We placed a closed petrol bucket at a distance of 20 cm from the sensor probe, collected measurements of each sensor, and then draw the measured curves.

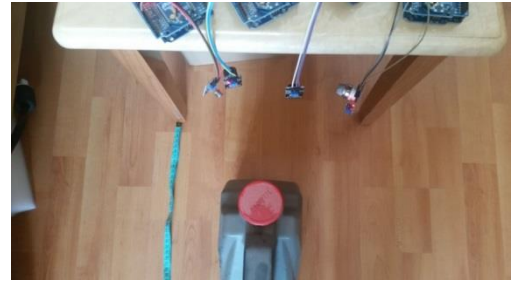

Figure 1 Sensor sensitivity test

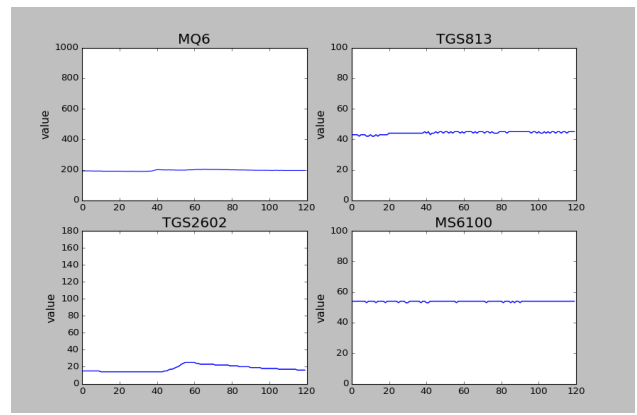

Figure 2 Measured response of different sensors

From Figure 2, we can see that the TGS2602 sensor has the best detection capability in the case of slight gasoline leakage in the air. Therefore, we choose to use TGS2602 in the design and system experiment of this paper.

## 3.2 Sensor deployment

The gas detection mechanism proposed in this paper aims to discover the abnormal gasoline vapor diffusion in the small space as soon as possible. The measurements of gas sensors depend on the gas concentration in the nearby air, and gas diffusion law determines the distribution of gas concentrations in different spatial positions.

According to the leakage gas density and the type of leakage source, the gas diffusion mode can be divided as mass diffusion and plume diffusion. When there is a large amount of sudden leakage, and the gas density is obviously higher than air, the gas diffusion shows a mass diffusion. In the mass diffusion, the gas rapidly forms smoke cloud in the air, and then moves with the air turbulence without diffusion or slow diffusion. When the gas density is close to the air, or in a very short period of time, the proliferation of its way is smoke plume diffusion. In most cases, a continuous leakage source is easy to form smoke plume diffusion.

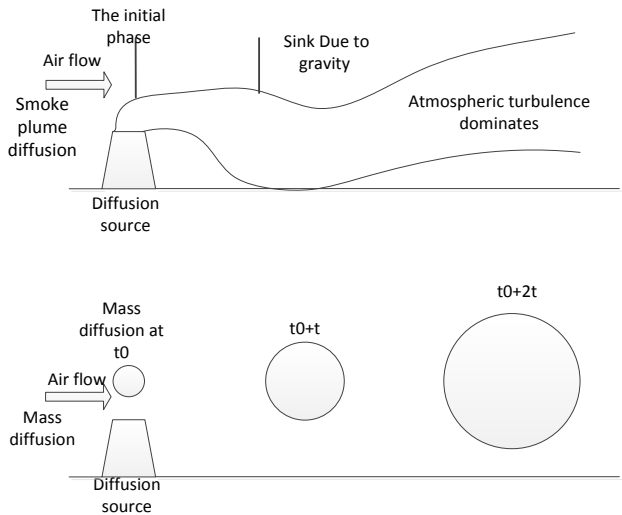

Figure 3 Smoke plume diffusion and mass dffusion model

For petrol gas leak diffusion, when the gasoline is stored in a closed gasoline barrel, or when the petrol barrel is open without shaking, the gas diffusion is dominated by smoke plume diffusion. In contrast, when the gasoline suddenly exposed to the air, much gasoline gas will spread in short time, and lead to a hybrid diffusion process of mass diffusion and smoke plume diffusion. The Figure 3 shows two examples of smoke plume diffusion as well as mass diffusion respectively. We can see that, under the smoke plume diffusion model, the gas concentration has a relatively stable spatial distribution. So gas diffusion behavior can be conjectured by analyzing correlations between measurements of the sensors located at different positions. Under the mass diffusion model, gas concentration at specific location is time-varying. So it needs to jointly analyze measurements collected at different times on one sensor to accurately determine the gas diffusion.

In order to verify the above inferences, we carried out a gasoline diffusion detection experiment in a small space. In this experiment, the gas detection sensors are deployed in the 0cm, 50cm, 100cm and 150cm heights, the distance between the sensor and the petrol bucket on the ground is 100cm. In the indoor environment, we open the petrol barrel, and shake the petrol barrel, and then draw the measurement curves of sensors respectively.

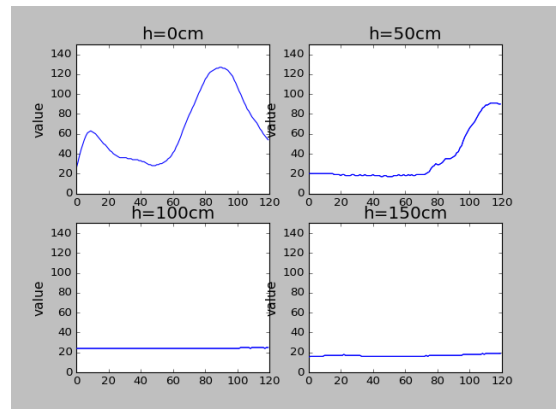

Figure 4 Sensor measurement curves with the barrel lid open

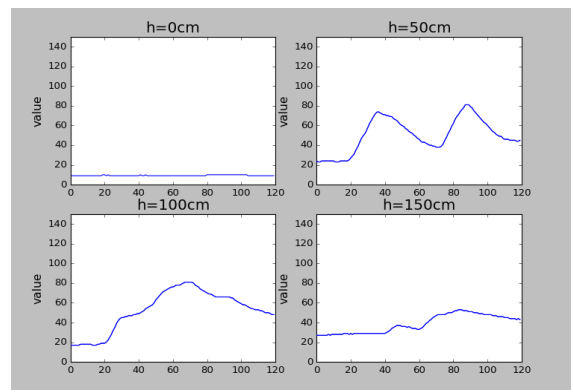

Figure 5 Sensor measurement curves with the barrel lid closed

As the petrol barrel is 30cm tall, we can infer that, from the above experimental results, in the volatile process (as shown in Figure 4), the gasoline vapor is first slightly elevated, and then gradually sinks. We can also see from the above experiment that, when shaking the oil barrel at the same position (as shown in Figure 5), compared to the natural diffusion process, it is more likely to cause the sensors deployed at higher places response to the diffusion process. These two conclusions are consistent with aforementioned gas diffusion model. Since sensors deployed at the height of 50cm (which can be considered common gasoline barrels and other high) is always more sensitive to detect the gasoline vapor, and in the long run the gasoline vapor will finally sink to the floor, we choose to deploy sensors at 50 cm height and on the ground (0 cm height) at the same time.

## 3.3 Experiment on Gasoline Diffusion Detection using Sensor Array

To further explore the ability of multiple sensors to detect gasoline diffusion process, we deployed a layered sensor array at a height of 50cm, and on the ground. As shown in the Figure 6, in this experiment, we have deployed 3 groups of 6 sensor nodes. The distance between two adjacent groups is 100cm. In the experiment, we control the air flowing from right to left by opening doors and windows in the room.

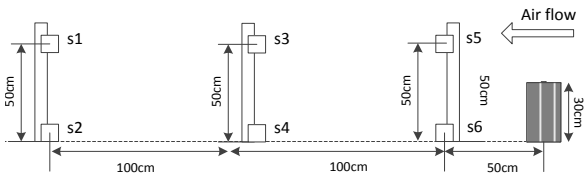

Figure 6 Detection of gas diffusion process

In this experiment, we collected measurements of the sensor groups after opening the petrol barrel, in case there is absence of air flow and in case there is air flow, respectively. The measurements are plotted in Figure 7 and Figure 8.

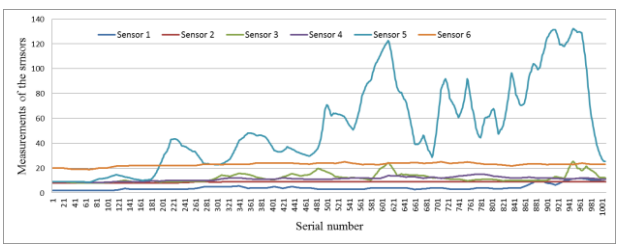

Figure 7 Measurement curves of sensors in an enclosed space

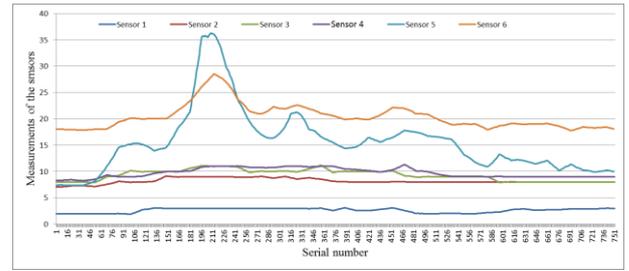

Figure 8 Measurement curves of sensors in an open space

From the Figure 7 we can see that, when the air circulation is poor, due to continuous evaporation of gasoline gas in the air, the concentration of gasoline gas continues rises on the whole. At the same time, due to the air turbulence, the relative gas concentration showed periodic characteristics. Whereas when the air circulation is good, in the early diffusion stage, the relative gas concentration will significantly increase due to mass diffusion, and then continued to decrease over time. At last, the smoke plume diffusion dominates the diffusion process, and each sensor node presents a stable relative concentration distribution according to its position.

Based on the above experimental results, we can conclude the typical characteristics of sensor measurements for combustible gas diffusion detection as periodicity, correlation and localness: 1) Due to the turbulence of the air, the measurement time series show fluctuations according to time. 2) Under the smoke plume diffusion model, the sensors near to each other show certain correlation in the measurements, especially for those deployed between upper and lower layers. 3) Fluctuations in sensor measurements due to gasoline vapor diffusions are generally limited to a small range without a consistent change in global sensor nodes.

## 4 Detection logic and system framework

Since in this paper we aim at the small space surveillance, target area size is generally within one-hop transmission range of the ZigBee radio, the remainder of this paper will be discussed based on the star network topology. For routing communication mechanism that may be used in larger networks will not be discussed in this paper. As a typical star network topology, our system consists of several battery-powered sensor nodes, and one power-supplied sink node. Sensor nodes perform environment sensing, local data processing and transmission, the sink node jointly analyzes the reporting data from multiple sensor nodes, determine the type of abnormal events, and finally alarm.

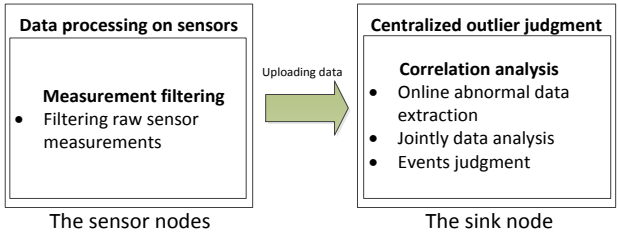

Figure 9 Data processing flow

The Figure 9 shows the data collection and processing performed at the sensor nodes and at the sink node, respectively in this paper. The overall data flow is divided into two phases: The sensor nodes collect the gas concentration information in the environment, after filter the noises in the sampling data, send them to the sink node. The sink node maintains multiple threads for every sensor node, performs online outlier detection for time series of each sensor node, and then jointly analyzes measurements from multi sensor nodes to judge outlier events. The specific data collection, processing, transmission, and analysis processes will be described in the next section

# 5 Data collection, transmission and processing

Based on the logic of combustible gas anomalous diffusion detection proposed in Section 3, this section will introduce the specific techniques we adopted at each stage. Since the system is a modular design, the realization of each module can also be extended in the future work in order to achieve further optimization.

## 5.1 Data Filtering on Sensor Nodes

In order to filter the high frequency noise, we use the moving average filtering algorithm to preprocess the measured time series of the sensor, and only send the filtered data to the sink node for further processing.

The moving average algorithm is implemented as follows:

a) Maintaining a time window of length N on each sensor node;

b) When new measurement arrives, it replaces the measurement on the tail of the time window, and then the arithmetic mean of the N measurements in the time window is took as the sampled value after filtering;

c) The value of N depends on the specific sampling interval.

## 5.2 Analysis and Extraction of Abnormal Data

In this paper, we use high-precision semiconductor combustible gas sensors to detect combustible gas composition in small space. In order to constrain the implementation cost of the system, this system will not include the complicated sensor calibration procedure. But on the other hand, the external temperature and humidity fluctuations, and even the sensor's instable power supply will affect the measurements of sensors. So in this system, after the preliminary data processing on the sensor nodes, the sink node need to further analyze the collected data flow. By distinguishing the cause of abnormal data changes, it is able to determine whether there is abnormal gas diffusion in the current monitoring region.

In this paper, we will use the aforementioned periodicity, correlation and localness characteristics (as proposed in section 3.3) to determine whether there is an exception event by analyzing the temporal and spatial correlation between measurements of different sensors deployed in the monitoring region. We analyze the correlation of different measurements by quantifying the time series similarity. For more accurate quantization calculation, it is first needed to extract the time series sub-segment from the original time series for matching. In this paper, we use time-series segmentation through extracting important points.

In the field of data mining, the local minimum and maximum points of time series are often used as the key points to divide the time series. Considering the periodic characteristics of volatile gas concentration variation, we want to choose the complete measurement fluctuation period series for analysis. Therefore, this paper will use the local extreme points of time series as the key point.

Suppose the measurement time-series of sensor X is represented as $X = \{x(t_i)\}_{i=1}^{n}$, wherein $[x_i, \ldots, x_{i+a}]$ is a subset of the time series. If there is a minimum value $x_{min}$ or a maximum value $x_{max}$ in the subset, then these two elements are referred as local minimum and maximum points.

The local maximum/minimum point extraction algorithm is as follows:

Algorithm 1: Extreme point extraction algorithm for time series

**Input**: Time interval a, measurement series X
**Output**: Local extreme points set
**Function**: Find out the local extreme points in X
**Algorithm**:
(1) $x_{min}=x1, x_{max}=x1$
(2) for j in 1 to m :
(3)    for i in j to j+a :
(4)       if $x_i > x_{max}$
(5)          $x_{max}=x_i, t_{max}=t_i$
(6)       else if $x_i < x_{min}$
(7)          $x_{min}=x_i, t_{min}=t_i$
(8)       end if
(9)    end for
(10) return $(x_{min},t_i),(x_{max},t_i)$

Through Algorithm 1, we can extract the maximum and minimum points to form a new time series $y(t_i)\ i \in \{1, \cdots, n\}$, $t_i \in (1, \ldots, m)$, wherein $N$ is the number of the original sequence elements, $m$ is the number of important points, and $a$ is a configurable parameter to control the number of output extreme points. Through the Algorithm 1, we can obtain some important extreme points in the time series, but when the curve changes smoothly, many redundant maximum and minimum points are included in the result extreme points set. In order to reduce the redundant extreme points, we have carried on further optimization based on the trend of the time series.

Considering that the rising and falling trend of sensor measurements is the most important feature in anomaly detection, we will remove the time series points which have less influence on the trend from the output extreme points set of the Algorithm 1. This process is completed by the Algorithm 2.

Algorithm 2 Trend optimization algorithm for time series

**Input**: Local extreme point time series X, trend tolerance $w$
**Output**: The optimized time series Y and the corresponding time stamp series T, the trend slope series K
**Functionality**: Filter redundant extreme points based on trends
**Algorithm**:
(1) for j in 1 to n:
(2)    i=1
(3)    $s_j = (x_{j+1}-x_j)/(t_{j+1}-t_j)$
(4)    if $s_{j+1}/s_j > w$ or $s_{j+1}/s_j < 1/w$
(5)       $y_i=x_j, t_i=t_j, k_i=s_j$
(6)       i=i+1
(7)    end if
(8) end for
(9) return $y_i, t_i$

From Algorithm 2, after the trend optimization, we can get the key points set. Then the sink node can divide the time series into several sub-segments according to output key points. Since during the gas diffusion process, the gas concentration fluctuates according to time, we need to intercept the complete trend field sub-segments to analyze the similarity of sensor measurements. Therefore, the starting point corresponding to the rising trend is taken as the segmentation point of each sub-segment. Suppose that the time-stamp set corresponding to the time-series sub-segmentation point is $T = \{t_1, t_2, \cdots, t_l\}$, we store it as a delimited label for the time series stored on the sink node.

After acquiring the set of trend slopes corresponding to each key point, i.e., $K = \{k_1, k_2, \cdots, k_l\}$, we can determine whether the subsequence is of exceptions based on $|k_i|$. Only when $|k_i| > k_{th}$, the sub-sequence is considered to be an anomaly sub-sequence, wherein $k_{th}$ is the threshold value.

## 5.3 Quantify the Similarity of Time Series

Although the sensor nodes with different distances from the diffusion source have different fluctuation amplitudes in the measured gas concentration curves, they have similar change patterns. For this feature, we choose to use the dynamic time warping distance (DTW) to quantify the morphological similarity between the various measurement sequences.

The DTW distance was first introduced by Berndt and Clifford in the mid-1990s to time series mining fields, and has been successfully applied in a lot of applications [12]. The DTW allows the time series to curve over the time axis. Unlike the Euclidean distance, the DTW is not exactly a point-to-point calculation, but rather it can skip several points in the matching sequence, so that the two sequences can be matched in a more "close-up" way. The DTW preserves the match between most of the pairs of points, while avoiding the shortcomings of Euclidean distance.

Suppose there are two time series of length m, n respectively:

$$q[1:m] = \{q_1, q_2, \cdots, q_m\} \quad (1)$$

$$c[1:n] = \{c_1, c_2, \cdots, c_n\} \quad (2)$$

The calculation of DTW distance for $q$, $c$ can be performed by constructing a distance matrix of size $m \times n$, called the bending matrix, as shown in Figure 10.

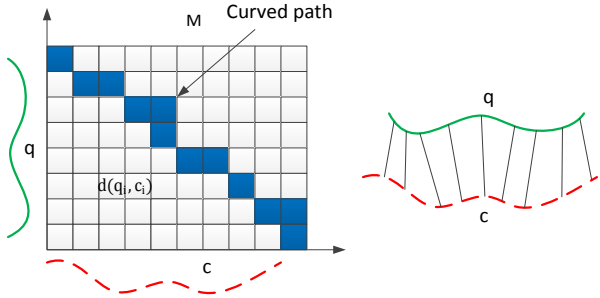

Figure 10 Bending matrix to calculate DTW

After constructing the bending matrix, the matching between $q$ and $c$ points is transformed into a curved path in the bending matrix from the square (1, 1) to the square (m, n): $W = w_1, w_2, \cdots w_l(max(m,n) \leq l \leq m + n)$. Define a mapping function $f_w: (q, c) \rightarrow W$ that maps the $(q, c)$ point pairs in the curved path to a square in the curved path:

$$w_k = f_w(q_i, c_j), 1 \leq i \leq m, 1 \leq j \leq n, 1 \leq k \leq l \quad (3)$$

So the DTW distance between q and c is transformed into figure out a curved path having the smallest distance in the bending matrix $W^*$:

$$DTW(q, c) = \underset{w}{argmin}(\sum_{i=1}^{c} w_i) \quad (4)$$

The equation (4) can be solved recursively according to equation (5):

$$\begin{cases} D_{dtw}(<>, <>) = 0 \\ D_{dtw}(X, <>) = D_{dtw}(<>, Y) = \infty \\ D_{dtw}(X, Y) = d(x_1, y_1) + min \begin{cases} D_{dtw}(X, Rest(Y)) \\ D_{dtw}(Rest(X), Y) \\ D_{dtw}(Rest(X), Rest(Y)) \end{cases} \\ d(x, y) = \|x - y\|_p \end{cases} \quad (5)$$

Wherein $Rest(X) = \{x_2, x_3, \cdots x_m\}$, $Rest(Y) = \{y_2, y_3, \cdots y_n\}$.

## 5.4 Analysis and Judgment of Abnormal Events

Based on the experimental results obtained in section 3.3, we present the judgment logic for abnormal gasoline vapor diffusion as in the Figure 11. When the two time series are normalized, they are considered to be correlated if their DTW distance is less than the specified threshold. When the time series data of a sensor node changes abnormally, the sink node will analyze whether the abnormal change has global correlation. If there is global correlation, we judge that the abnormal changes in the measurements are caused by the change in the external environment, and there is no abnormal gas diffusion. If there is no global correlation, the sink node will continue to judge whether the measured anomaly is time-dependent. If there is repeated abnormal fluctuation in the measurement node of a single measurement node, we judge that there is a combustible gas diffusion source exists near the sensor node. After performing the periodic judgment, local spatial correlation judgment is performed. If a sensor node and the nearby sensor nodes simultaneously appear abnormal fluctuation of the measured value, we judge that there is a combustible gas diffusion source nearby.

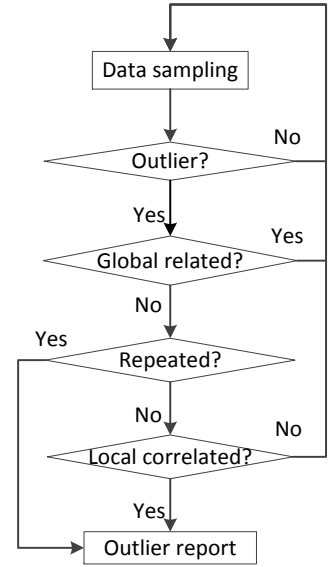

Figure 11 The flow chart of abnormal gas diffusion judgment

# 6 Experimental setup and performance evaluation

In order to verify the overall performance of the system, we set up a complete experimental environment testing system to detect different types of gasoline vapor diffusion in this section. The Figure 12 Experimental environment shows the experimental scenario. In the experiment, a total of 12 sensor nodes were deployed, which were divided into two layers. The first layer is located on the ground, including sensors A0, B0, C0, D0, E0 and F0. The second layer is at a height of 50 cm, including sensors A1, B1, C1, D1, E1 and F1. The distance between the sensors was 2 m in the x-axis direction and 2 m in the y-axis direction. In the figure, two experimental positions are indicated. In position 1, the diffusion source (gasoline drum) is 1 m from both the x-axis and the y-axis. At position 2, the diffusion source is 1 m from the node C0 and the node D0. In the experiment, the diffusion source is a 5L petrol barrel with a 10cm diameter lid and 35cm height.

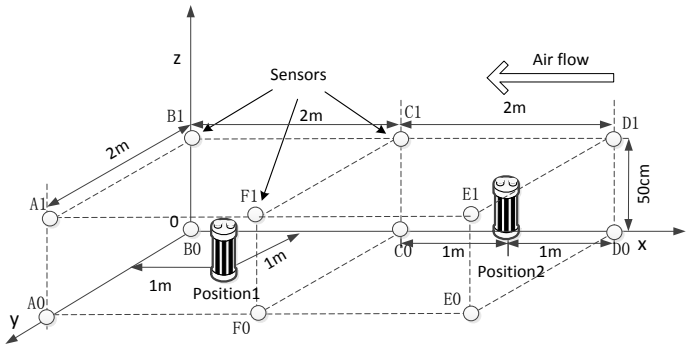

Figure 12 Experimental environment

In this experiment, we first test the sensor network's detection capabilities on the normal gasoline diffusion process, and then test the sensor network's anti-false alarm capability for typical external events. The experiment is divided into 6 parts. The first 4 parts are the detection of closed oil drum in open space, the detection of closed oil drum in closed space, the detection of open oil drum in open space, the detection of open oil drum in closed space. In the fifth part, an anti-false alarm detection is carried out. We artificially make the air circulate, change the ambient temperature, and shake sensor nodes to force measurements vary. In all of the above tests, the basic sensor temperature compensations are included.

For the first four groups of diffusion source detection experiments, we placed the petrol barrel at positions 1 and 2, and opened the barrel lid immediately in the latter two experiments, recorded the alarm time of the sysem to above operations. The process of each experiment lasted 2 minutes, and repeated 20 times. Figure 13 ~ Figure 16 shows the cumulative distribution of the alarm delay of the first four groups of experiments. In these graphs, we treat the time out reports as failure detections.

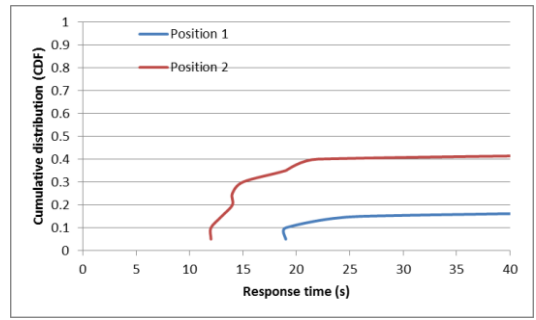

Figure 13 Closed gasoline barrel in the open space

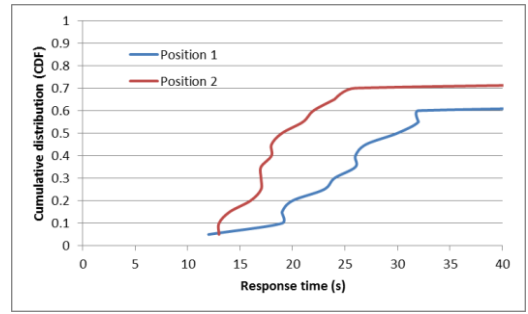

Figure 14 Closed gasoline barrel in the enclosed space

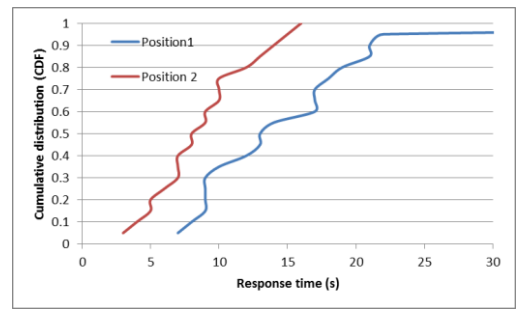

Figure 15 Open gasoline barrel in the open space

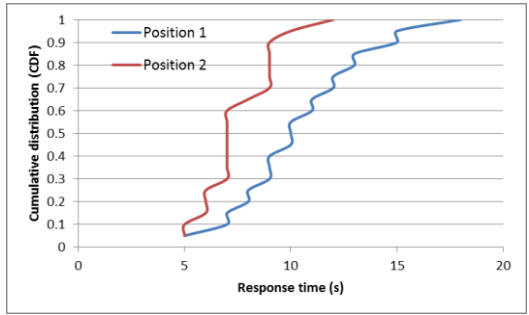

Figure 16 Open gasoline barrel in the enclosed space

From the above experimental results, it can be seen that, when the gasoline barrel is at position 1, the system has a higher detection delay than that of position 2. This is because when the barrel is at position 1, the diffusion source is farther away from the sensor nodes, and the changes in the sensor measurements become weaker, requiring more data to determine the causes of the abnormal events. For the first two groups of experiments with sealed petrol barrels, small amount of gas leakage can lead to a significant reduction in event detection rates. Especially in the first group of experiments, when we place the closed petrol barrel in the open space, the detection ratio always lies under 40%. We can also see that, the detection rate will significantly increase when the distance between diffusion source and the monitoring sensors decreases. As in the first experiment, the system had a 25% improvement in detection rate when we move the gasoiline barrel from position 2 to position 1, and also about 10% improvement in the second experiment.

In the fifth group of experiments, we test the sensor network system's fault tolerance for external interference events. At first, we forced the air flow using the fan, then we open the air conditioning to change the ambient temperature, and finally we shake the sensor node to make disturbance. In the above actions, each action is repeated 20 times, and then we record the alarm times of the system as shown in Table 1.

Table 1 System alarm performance

| Events | Correct alarm times | False alarm times | False alarm rate |
|---|---|---|---|
| Force air flow | 19 | 1 | 5% |
| Change Ambient Temperature | 20 | 0 | 0% |
| Shake sensors | 17 | 3 | 15% |
| **Total** | **56** | **4** | **6.7%** |

In the case of forced air circulation, the system has one false judgement in 20 attempts, the false alarm rate is 5%. In the case of forced ambient temperature changes, all the judgements are made correctly. In the case of artificially shaking sensor nodes, the system has 3 false judgements in 20 attempts, the false alarm rate is 15%. In this experiment, since we are using very tense external interference, especially the air flow intensity and the amplitude of artificial sensors are higher than the general situations, the overall false alarm rate is controlled under 7%, indicating good anti-jamming capability.

## 7 Conclusion

In this paper, a combustible organic liquid detection framework is proposed to detect fire hazard in small public places based on sensor network. As most flammable organic liquids have obvious volatilization characteristics, this paper realizes the detection of fire through measuring the concentration of volatile organic compounds in the air. For the most common gasoline gas detection, this paper compares and analyzes the detection response of common combustible gas sensors to petrol vapor, and selects the sensor with the best detection sensitivity as the basic data acquisition module. Based on the gas diffusion model and experiments under the typical small space scenario, we also proposed sensor deployment scheme, as well as the judgment logic for abnormal gasoline vapor diffusion.

After determining the gas detection framework and data flow analysis logic, this paper designs concrete implementation schemes in every data collecting and processing section. The schemes include filtering noisy sensor samples on the sensor nodes using the moving average filter algorithm, segmenting time series on the sink node, and time series segment matching based on DTW distance quantization.

In addition, we also deployed a sensor network in the real indoor environment to verify proposed monitoring framework and corresponding data processing mechanisms. The experimental results show that, our system can detect the gasoline barrel in the room (small space) in short time in most cases, while maintaining very low false alarm rates to typical external interferences.